\documentclass[journal]{IEEEtran}

\usepackage{fullpage}
\usepackage{setspace}
\usepackage{amssymb}
\usepackage{amsmath}
\usepackage{epsfig}
\usepackage{graphicx}
\usepackage{algorithm}
\usepackage{algorithmic}
\usepackage{bm}
\usepackage{color}
\usepackage{cite}
\usepackage[normalem]{ulem}
\usepackage{array}
\usepackage{enumerate}
\usepackage{float}

\newtheorem{definit}{Definition}

\newcommand{\ie}{\emph{i.e.}}
\newcommand{\eg}{\emph{e.g.}}

\begin{document}

\title{Inband device-to-device relays in cellular networks}

\author{Bilal Sadiq, Saurabh Tavildar, Junyi Li\\
        Qualcomm Research, Bridgewater, NJ, USA\\
        \{bsadiq, tavildar, junyil\}@qti.qualcomm.com\\
        \vspace{0.15in}
        November 18, 2013
\thanks{Drafted May 14, 2013; revised November 18, 2013.}}
\maketitle

\begin{abstract}
	A new design for two-hop opportunistic relaying in cellular networks is proposed, with the objective of throughput improvement. We propose using idle UEs in a cellular system with better channel to the base station to relay traffic for active UEs. One of the key ideas proposed is the use of (only) uplink spectrum for the Access links, and corresponding interference management schemes for managing interference between Access and Backhaul links. The proposed algorithms and architecture show a median throughput gain of $110\%$ for downlink and $40\%$ for uplink in system simulations performed according to the 3GPP methodology.
\end{abstract}


\section{Introduction}
\label{sec:intro}
We propose a design for two-hop decode-and-forward relaying in cellular networks, primarily for throughput improvement in the low and median throughput regime, on both the downlink (DL) and the uplink (UL) paths.   The Backhaul between the base station and relaying mobile device -- the ``longer'' hop -- is the conventional cellular wide area network (WAN) link, whereas the Access link between the relaying and relayed devices -- the ``shorter'' hop -- is an inband device to device (D2D) link. Inband Access links reuse (parts of) the same band that the WAN operates in. D2D discovery and communication are currently being studied in the Standards \cite{3GPP22803}.

\begin{figure}
\centering
\scalebox{1}{\includegraphics[trim = 0mm 7mm 0mm 2mm, clip, width=3.2in]{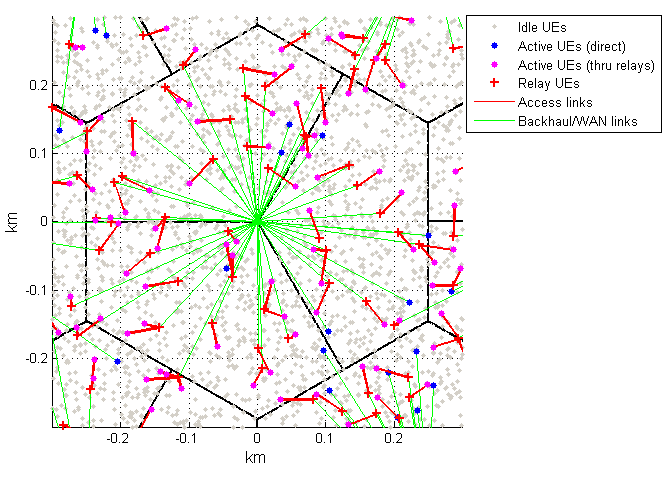}}
\vspace{-0.1in}
\caption{System model. Active UEs connect to eNB through relay UEs in their vicinity, or directly if no nearby UE is suitable for relaying. Access link between active and relaying UEs is a D2D link, whereas backhaul between relaying UE and eNB is the conventional cellular link.}
\label{fig:sysModel}
\vspace{-0.2in}
\end{figure}

The network topology is depicted in Fig.~\ref{fig:sysModel}.  One of the main motivations for the proposed two-hop architecture is to leverage large number of UEs that are part of a cellular network but are idle most of the time, and can be co-opted to improve overall system performance by relaying. However this poses challenges in terms of defining new interference management, relay association, and UE power management techniques.  In this paper, we address these challenges and show the gains of the proposed architecture through system simulations using methodology adopted in 3GPP \cite{3GPP36814}.


Layer 3 Relay Nodes were standardized in LTE Release 10 \cite[Sec.~4.7]{3GPP36300Rel10} and further enhancements are under consideration as part of wider heterogeneous network study.  These Relay Nodes however are a fusion of a scaled-down lower power base station (eNB) and an ordinary UE: they appear as eNB to the UEs being relayed, and they mimic some of the UE functions on the Backhaul to the Donor eNB.  Due to cost and power, Relay Nodes are not envisioned as add-on to ordinary UEs but are designed as stand-alone devices.  Moreover, since inband relays create further edges in the network, their deployment typically requires site planning \cite{BulakciSalehHama12} and advanced interference coordination schemes \cite{DamnjanovicMontojoTingfang11}.  Both these factors limit the number of Relay Nodes when compared with the ubiquity of UEs.  The proposed design exploits that ubiquity and repurposes UEs as relays, thus creating nearly one-to-one ratio between active devices and relays.
The most important gains of the proposed design are summarized in the following:
\begin{itemize}
    \item {\bf Exploiting shadow-diversity through relaying:} a UE's throughput to a large extent depends on the shadowing and pathloss to the serving base station. However, the shadowing is known to be uncorrelated over short distances of few tens of meters \cite[Sec.~B.1.2.1.1]{3GPP36814}\cite{Gudmundson91}. We exploit this fact to find UEs to relay that are at a short distance but have much better geometry to the serving base station.  We call this shadow-diverse relaying. In section \ref{sec:caseOppRly} we show that shadow-diverse relaying can produce over $10$ dB improvement in median SINR, which leads to a $200$\% increase in median throughput, and in section \ref{sec:implementation} we propose power efficient protocols for finding and associating with such a relay.
    \item {\bf Uplink spectrum for Downlink traffic:} in a typical (FDD) cellular system, DL spectrum is more congested than UL spectrum -- motivated by this and constrained by certain regulatory limitations on UEs transmitting in DL spectrum, we propose that the Access links use the UL spectrum for {\em both} directions. Other benefits of reusing UL spectrum, in particular simpler interference management, are discussed in Section \ref{sec:caseUlSpect}.
    \item {\bf Access$\leftrightarrow$UL interference management:} in a two hop relay architecture, interference management is needed across Backhaul and Access links as well as among Access links themselves.  However, typically the longer Backhaul link proves to be the throughput bottleneck -- motivated by this we propose an underlay approach for Access links by explicitly managing their interference to the base station (and hence the Backhaul link) via tight power control to the base station, and with minimal signaling overhead to the system.  Note that the areas of low throughput (or worse pathloss to base station) are naturally conducive to the reuse of UL resources.  Note also that a power control based solution for Access-to-WAN interference management is feasible only because of the proposed architecture of using UL spectrum for Access links; a similar approach would be harder to achieve on the DL spectrum.  Lastly, given the spatial and temporal (due to scheduling) sparsity of UL transmissions, the UL to Access interference can be left unmanaged, as shown in  Section \ref{sec:caseUlSpect} (Fig.~\ref{fig:spatialSigPow}).
\end{itemize}

The rest of the paper is organized as follows.
We demonstrate potential gains of  shadow-diverse relaying in Section \ref{sec:caseOppRly}.
We justify the reuse of UL spectrum for the Access link and introduce some of the key interference management ideas in \ref{sec:caseUlSpect}.
We provide a detailed design for relay association as well as interference management, and provide system simulation results in \ref{sec:propDesign}.
A brief discussion on implementation and technology aspects is presented in Section \ref{sec:implementation} before concluding the paper in Section \ref{sec:conclusion}.

\section{A case for shadow-diverse relaying}
\label{sec:caseOppRly}

\begin{figure}
\centering
\scalebox{1}{\includegraphics[trim = 0mm 0mm 0mm 0mm, clip, width=3.1in]{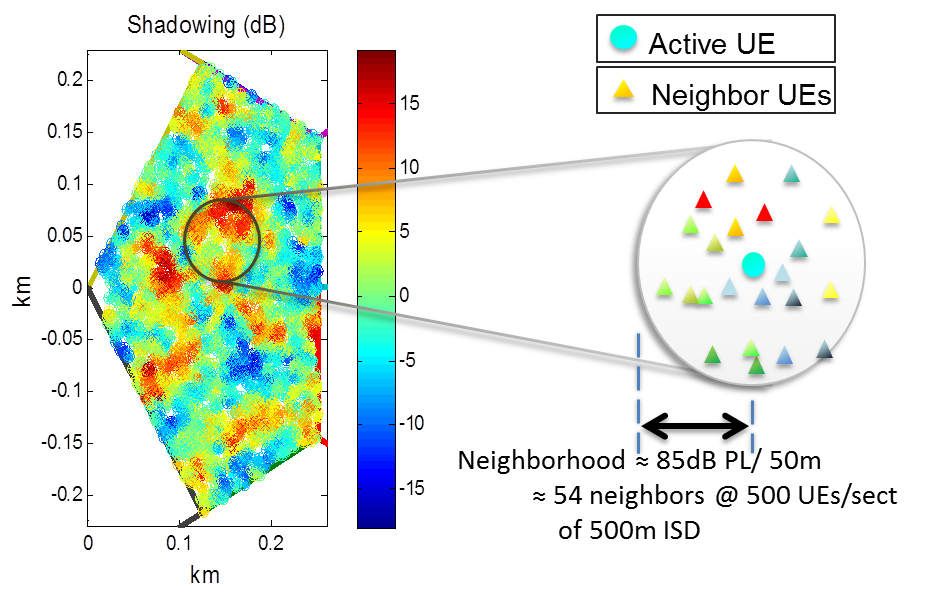}}
\caption{Harvesting multiuser diversity of the nearby idle devices, at the scale of shadowing. }
\label{fig:concept}
\end{figure}

The ubiquity of (D2D-enabled) user devices that can also act as relays -- even if over only few tens of meters -- can provide an advantage akin to the user walking outside a building for better reception. To demonstrate this quantitatively, we look at the nature of spatial shadowing correlation as well as its impact on DL and UL SINRs.  It was experimentally shown \cite{Gudmundson91} that shadowing (on dB scale) can be modeled as spatially correlated Gaussian random field, with the normalized autocorrelation between two points at distance $\Delta d$ given by,
$$
	R(\Delta d) = \exp(-\frac{\Delta d}{d_{corr}}),
$$
where the correlation distance $d_{corr}$ depends on the environment.  Indeed 3GPP has adopted the same model for LTE  evaluation \cite[Sec.~B.1.2.1.1]{3GPP36814} with $d_{corr}$ ranging from 6m to 50m for various urban and indoor environments.  Moreover, at any point, the shadowing values with respect to different base stations have cross-correlation of $\frac{1}{2}$. Fig.~\ref{fig:concept} illustrates spatial correlation of shadowing for $d_{corr}=25m$ (and $\sigma=7dB$). Other details of the WAN and D2D channel models are given in Appendix.

\begin{figure*}
\centering
\scalebox{1}{\includegraphics[trim = 0mm 0mm 5mm 0mm, clip, width=6.4in]{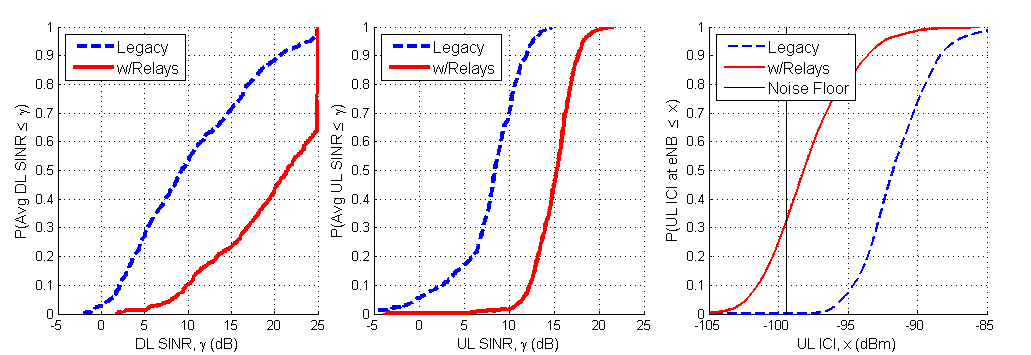}}
\caption{SINR and ICI CDFs.  Improvement in average (left) DL and (middle) UL SINR if all active devices were replaced by their relays. Improvement in UL predominantly comes from the (right) reduction in ICI.}
\label{fig:upperBound}
\end{figure*}

Next we quantify the potential gains of using a nearby device with better DL SINR as a {\em relay} between the base station and the {\em edge} device\footnotemark. \footnotetext{The term `edge' device will be used to refer to the relayed UE, \ie, the UE at the edge of the two hop link.}
Fig.~\ref{fig:upperBound} shows the improvements in average DL and UL SINRs {\em if the active devices were replaced by their respective neighbor with the best DL SINR.}  By `neighbor' we mean any device with $\leq$85dB D2D pathloss to the active device (or equivalently, within about 50m distance).   Key observation is that {\em few tens of meters is a sufficient relay search radius to exploit spatial variability of shadowing}, improving median DL SINR by 13dB and UL by 7dB.  Note the improvement in UL SINR is predominantly due to the reduction in inter-cell interference (ICI), as shown in  Fig.~\ref{fig:upperBound}(right), and to a lesser extent, due to slightly stronger received signal at the serving eNB when the selected neighbor has better pathloss than the original active device.  Indeed, a device can use a different criterion (than DL SINR) for selecting an UL relay, \eg, the ratio of serving to interfering WAN channel pathloss, or an added requirement that the pathloss of relay to the serving base station be better than that of the edge device itself.  Opting for simplicity we defer such optimizations to future work. An analytical approximation of shadowing-diversity offered by nearby devices, for a specific relay selection scheme, is also provided in \cite{CalcevBonta09}.

\section{A case for using the UL spectrum for Access links}
\label{sec:caseUlSpect}

In addition to regulatory restrictions on UEs transmitting in DL spectrum, there is also a technical case for using the UL spectrum, built on two conjugate observations: that existing interference in UL spectrum is naturally conducive to spatial reuse by short Access links, and that the interference to base station introduced by Access links is easily managed through power control to the base station.

\subsection{WAN-to-Access-link interference}
\label{subsec:wan2accIM}
\begin{figure*}[!]
\centering
\scalebox{1}{\includegraphics[trim = 7mm 0mm 40mm 5mm, clip, width=6.4in]{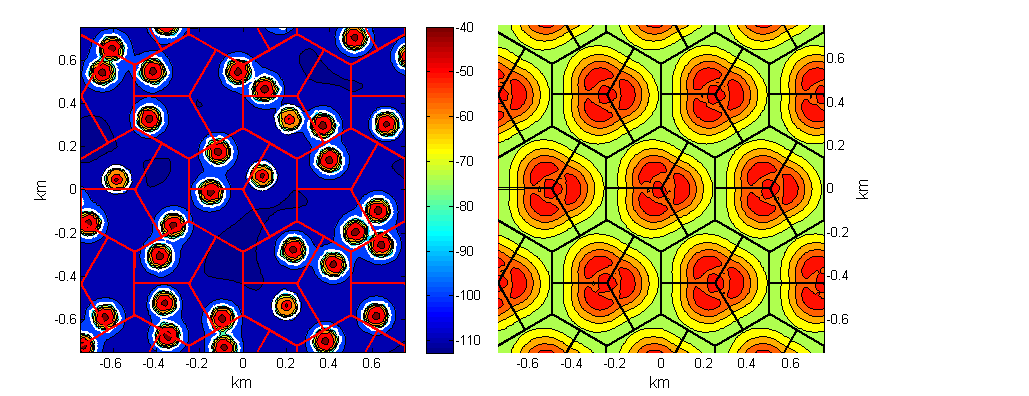}}
\caption{Snapshot of total signal power (left) in UL spectrum; -90dBm contour curve is plotted in white for reference.  One UE per sector is scheduled and allocated the entire 10MHz bandwidth. (Right) That in  DL spectrum.}
\label{fig:spatialSigPow}
\end{figure*}

Let us first consider the WAN-to-Access-link interference.  In UL spectrum the sources of interference are weaker UE transmissions arriving over a weaker D2D channel; by contrast, the interference in DL spectrum would originate from higher power base station transmissions arriving over a stronger WAN channel (due to antenna placement).  For comparison, Fig.~\ref{fig:spatialSigPow} shows a snapshot, in an arbitrary subframe, of spatial distribution of power in the UL vs the DL spectrum, as seen by a UE.   We note that in the UL spectrum, except for a few bright blobs centered at the UEs {\em scheduled for UL transmission in this subframe}, the UL spectrum is amenable to reuse in the remaining {\em darker} space.  As different UEs are scheduled from subframe to subframe, almost every location gets to see low interference for some fraction of time.

\begin{figure*}
\centering
\scalebox{1}{\includegraphics[trim = 5mm 0mm 10mm 0mm, clip, width=3.2in]{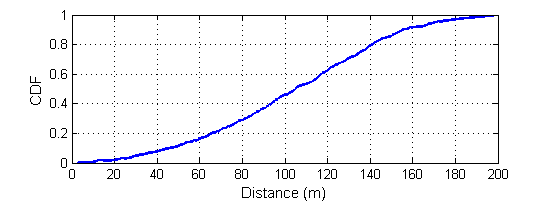}}
\scalebox{1}{\includegraphics[trim = 5mm 0mm 10mm 0mm, clip, width=3.2in]{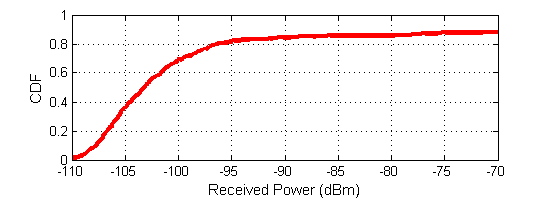}}
\caption{(Left) CDF of distance to the nearest UL transmitter on a given RB from point $\left(\mbox{ISD}/3,0\right)$. (Right) CDF of total received power in UL spectrum at point $\left(\mbox{ISD}/3,0\right)$; note $80^{th}$\%-tile is reasonably small (below -90 dBm).}
\label{fig:temporalSigPow}
\end{figure*}

To further illustrate this temporal aspect, Fig.~\ref{fig:temporalSigPow}(left) gives the CDF of the distance to the nearest UL transmitter, on an arbitrary resource block (RB), from reference point $\left(\frac{\mbox{ISD}}{3},0\right)$ in the standard 2-tier, 19 cell deployment with inter-site distance (ISD) of 500m.  We note that the mean is about $0.22\times\mbox{ISD}=110$ meters -- much larger than the Access link `length' needed to exploit shadowing diversity.   More specifically, Fig.~\ref{fig:temporalSigPow}(right) gives the CDF of total received power at reference point $\left(\frac{\mbox{ISD}}{3},0\right)$ from UL transmissions in the 19 cell deployment.  We note that even the $80^{th}$ percentile is reasonably small (below -90 dBm). {\em We conclude that, as-is, the UL spectrum is amenable to spatial reuse by Access links.}\footnotemark~ Next we consider the Access-link-to-WAN interference.

\footnotetext{Whereas, spatial reuse of DL spectrum for Access requires dominant interference cancelling receivers \cite{DamnjanovicMontojoJoonyoung12} and cell range expansion through handover biasing and adaptive resource partitioning \cite{DamnjanovicMontojoTingfang11}.}

\subsection{Access-link-to-WAN interference}
Just as all connected UEs are power controlled in UL to have nearly equal signal strength at the base station \cite[pp.~464-471]{SesTouBak09}, all Access link transmissions can also be power controlled accordingly to ensure that the interference (to base station) introduced by each link is at least, say, $20$dB below the UL signal strength.   That is to say, the Access link UEs can derive their transmission power (for Access link) from their UL transmit power with a 20dB backoff.  This backoff approach is a simplification of the scheme proposed in \cite{JanisYuDoppler09} for coexistence of D2D communication with cellular, and lends itself favorably to the relay use case: the {\em farther} the UE from base station (and thus more in need of a relay), the higher the allowed transmission power for Access link communication.  This underlay framework admits multiple optimizations such as:
\begin{itemize}
    \item The transmit power backoff parameter can be dynamically set and broadcast by the base station to measure and control total interference from Access links.
    \item With interference budget per Access link fixed, the total number of simultaneously active Access links can also be implicitly controlled by the rates allocated to relays on the Backhaul, as Access links naturally shuts down when the relay buffer is empty (DL) or full (UL).
    \item Lastly, base station can calculate its {\em interference price} and precisely decide if an Access link is too costly for the rate improvement it offers, and therefore schedule that edge UE directly without the relay.
\end{itemize}
Investigation of these optimizations, once again, is deferred to future work.

\subsection{Access-link-to-Access-link interference}
Finally, we comment on Access-link-to-Access-link interference.  Access links being short, low power and sparse, they can be allowed to reuse the entire UL spectrum and need not be actively scheduled by the base station (unlike WAN links which share the UL resources.)  However, as an implication of Birthday Problem \cite{Grimmett3rdEd}, it is likely that even in a sparse population a few Access links strongly interfere with each other and therefore prove ineffective without further interference coordination.  In Section \ref{sec:simResults} we present simulations with and without interference management (TDMA) among Access links, and note that most of throughput gain is realizable even without any interference management or sophisticated WAN scheduling.

\section{A candidate design for a D2D relay system}\label{sec:propDesign}
Having explored the extent of SINR improvements provided by shadow-diverse relaying (Section \ref{sec:caseOppRly}) and the feasibility of reusing UL spectrum for short Access links (Section \ref{sec:caseUlSpect}), we now layout a candidate system design and investigate its performance through simulation. In picking a candidate design to demonstrate the concept, we will opt for simplicity, deferring sophisticated optimizations to future work.

\begin{definit}
	{\em Relay candidate.} A device is a relay candidate of an active device if
	\begin{enumerate}[i.]
		\item the two devices are in the same sector, and
		\item pathloss between the two devices is smaller than $p^{max}_{acc}$.
	\end{enumerate}
\end{definit}

We limit the link budget for Access link $p^{max}_{acc}$ to a very conservative value of 85dB (\ie, about 50m link length).  Relay search over 50m yields significant opportunity to exploit shadowing diversity, as seen in Section \ref{sec:caseOppRly}.  Moreover, short Access link means the transmit power can be curtailed to enable an underlay approach, as well as to enable spatial reuse across Access links.  From amongst the relay candidates, a relay is selected as follows.

\begin{definit}\label{def:rlySelect}
	{\em Relay selection.} From amongst its relay candidates, an active device selects one that has the highest DL SINR, provided it is also higher than the active device's own DL SINR.
\end{definit}

A device that does not select a relay (because, \eg, the device itself has the best DL SINR amongst the candidates) naturally connects to the base station directly.  Moreover, multiple active devices may sometimes select a common relay, although this is rare given small search neighborhood and spatial sparsity of active devices (see Fig.~\ref{fig:sysModel}).

Instead of orthogonalizing WAN and Access links, we allow reuse 1 of UL resources on each Access link.  That is, each Access link is allowed be active at all times -- barring a natural half-duplex constraint described later -- in the entire UL spectrum (FDD), or in the entire spectrum during UL subframes (TDD).  However, the power for Access link transmission is tightly controlled to limit the interference to WAN, as follows.

\begin{definit}\label{def:txPower}
	{\em Access link transmit power.} Transmit power on the Access link by any device (edge or relay) is $P_{UL}-\Delta_{acc}$ dBm, where $\Delta_{acc}$ is a design parameter and $P_{UL}$ dBm is the UL transmit power of the device if the device were allocated the entire UL spectrum.
\end{definit}

We set the backoff parameter $\Delta_{acc}$ to $20$dB in subsequent simulations. Since the devices are power controlled on the UL, the above method of deriving Access link power from the UL power can be seen as allocating each device an interference budget that the device is allowed to cause to the WAN. It is worth noting that the Access link transmit power implicitly depends on the device's pathloss {\em to serving  base station}.  As a result, the edge UEs, which typically have worse pathloss to the base station than the relay UEs, tend to have higher transmit power allowance than the relay UEs. Therefore we expect to see edge-to-relay (\ie, UL) Access links to have better SINR than the relay-to-edge links.

A word regarding the half-duplex constraint on Access links:
\begin{definit}
    {\em Half-duplex constraint and relative priority of UL vs Access link.} A device is not allowed to be simultaneously active on the WAN UL and the Access link. Therefore, the Access link is not active in the subframes where the relay or the edge device are scheduled by the base station for an UL transmission.
\end{definit}

While the Access links are not directly scheduled by the base station, the `on' duration of an Access link implicitly depends on the rate being allocated to the relay on the (base station scheduled) Backhaul. More specifically,
\begin{definit}
    {\em Backhaul and Access link coupling through relay buffer.} Relay UEs are configured with a small (edge UE specific) buffer to hold the relayed traffic temporarily.  When the buffer is full, the link feeding the buffer shuts down.  That is,
    \begin{itemize}
        \item if the buffer is associated with UL traffic, then the Access link shuts down (since the UL relay buffer is fed over the Access link by the edge UE);
        \item if the buffer is assocated with DL traffic, then the base station stops scheduling relay Backhaul.
    \end{itemize}
    Similarly, when the buffer is empty, the link draining the buffer naturally shuts down, namely, the Access link for the DL traffic and Backhaul for the UL.
\end{definit}

Lastly, we describe a possible interference management scheme amongst Access links.
\begin{definit}
	{\em Interference management amongst Access links.}  Two Access links are said to interfere with each other if the (pairwise) SIR at any one link due to the interference from the other is below $\gamma_{acc}$.  Between any two interfering links in a given subframe, one link is chosen uniformly at random to yield to the other.  Therefore, an Access link transmits if it does not yield to any interfering links in that subframe.
\end{definit}

In short, interfering Access links are active in a time division multiple access (TDMA) fashion such that an Access link with $d$ interfering links is active about $\frac{1}{d+1}$ fraction of the time. This interference management/multiple access (MAC) approach is derived from the distributed FlashLinQ MAC algorithm; see \cite[Sec.~I-A]{WuTavildarShakkottai10} for motivation behind SIR-based yielding and \cite[Sec.~III]{WuTavildarShakkottai10} for details of distributed SIR computation and yielding.
Note that if SIR threshold $\gamma_{acc}$ is set to $-\infty$, no local interference management amongst Access links takes place (this is one of the simulated cases in Section \ref{sec:simResults}.)  The other case is $\gamma_{acc}=5$dB.

To recap, each Access link transmission spans entire UL spectrum and each Access link is always `on' except when (1) the relay buffer is empty (in case of DL) or full (in case of UL), (2) the relay or the edge device is scheduled for an UL transmission, or (3) the Access link has yielded to another Access link (assuming Access link interference management scheme is being used.)

\begin{figure*}
\centering
\scalebox{1}{\includegraphics[trim = 10mm 0mm 10mm 0mm, clip, width=6.4in]{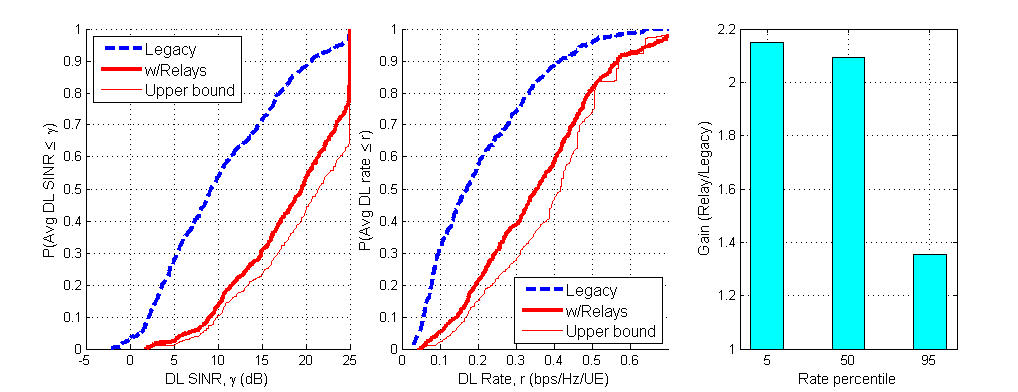}}
\caption{Improvement in DL SINR and throughput.}
\label{fig:dlwoIM}
\vspace{-0.1in}
\end{figure*}
\begin{figure*}
\centering
\scalebox{1}{\includegraphics[trim = 10mm 0mm 10mm 0mm, clip, width=6.4in]{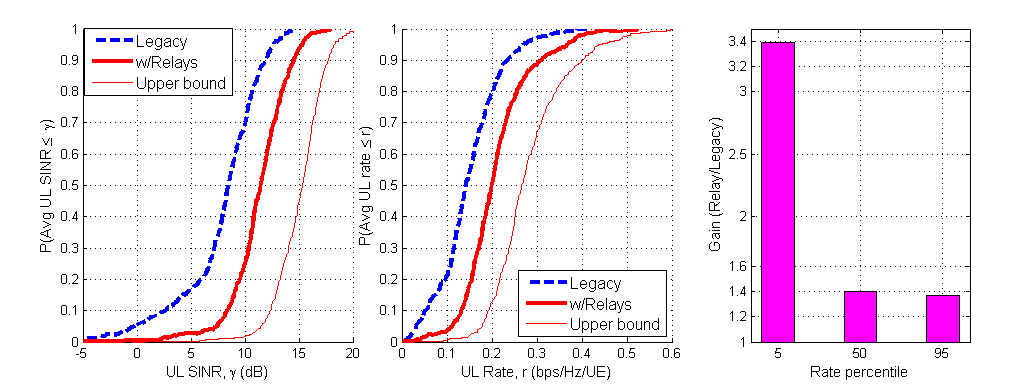}}
\caption{Improvement in UL SINR and throughput.}
\label{fig:ulwoIM}
\vspace{-0.1in}
\end{figure*}

\subsection{Simulation results and discussion}
\label{sec:simResults}
We simulate the above system under full-buffer traffic. Scheduling and resource allocation in the base station for both DL and UL are performed by a proportional fair (PF) scheduler.  In subframes where the buffer status of the relay does not permit scheduling the relay Backhaul, the proportional fair scheduler instead considers scheduling the UE directly.  Therefore, the end to end links that are Backhaul limited will have (nearly) all the data flowing through relay, whereas those that are Access link limited will have as much data flowing through relay as permitted by the Access link and the remaining share of the UE's data flowing directly to/from the edge UE.  Access links can be the bottle neck due to low transmit power budget (see Definition \ref{def:txPower}), strong interference from another nearby Access link, or being {\em surrounded} by many UL transmitting UEs. Relays that offer less than $5\%$ increase in rate over going directly are dropped and those UEs are scheduled directly by the base station.

A relay-buffer-and-channel-aware scheduler that opportunistically decides between the direct and through-relay path can yield higher throughput; such optimizations once again are deferred to future work.

Next we discuss the simulation results, first for the case without interference management (IM) amongst Access links and then for the case with IM.


\subsubsection{Without interference management amongst Access links}
\begin{figure}
\centering
\scalebox{1}{\includegraphics[trim = 0mm 0mm 0mm 0mm, clip, width=3.2in]{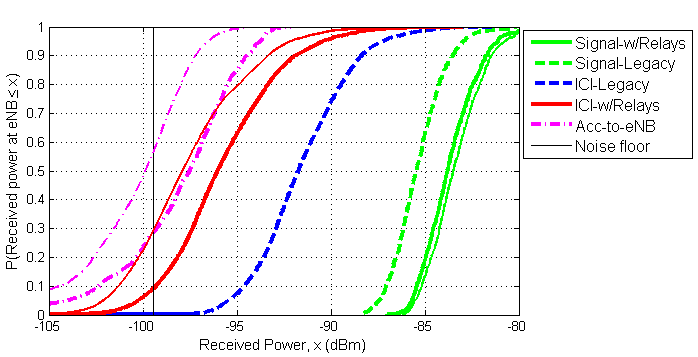}}
\caption{Various interferences and signal strengths at eNB, with ($\gamma_{acc}=5$dB) and without ($\gamma_{acc}=-\infty$) IM amongst Access links. Thin lines denote the case with IM.}
\label{fig:intf}
\end{figure}

\begin{figure}
\centering
\scalebox{1}{\includegraphics[trim = 0mm 0mm 0mm 0mm, clip, width=3.0in]{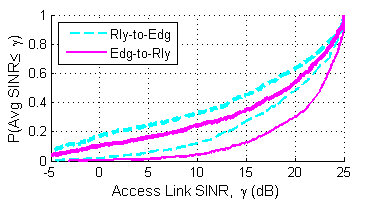}}
\caption{Access Link SINRs (forward and reverse). Thick curves are without IM amongst Access links, whereas the thin curves are with IM.}
\label{fig:acc}
\vspace{-0.2in}
\end{figure}

First we present results for the system without interference management (IM) amongst Access links, \ie, the case with $\gamma_{acc}=-\infty$.  DL-specific user throughput results are summarized in Fig.~\ref{fig:dlwoIM} and UL-specific in Fig.~\ref{fig:ulwoIM}.  We make the following observations:
\begin{enumerate}
  \item Average DL SINR improves significantly with the use of relays, with median improving by about $10$dB.  Note the average DL SINR of a UE is convex combination of the UE's and its relay's DL SINR, depending upon relatively how often each was scheduled.  In the low SINR regime, the CDF matches well with the upper bound obtained from Fig.~\ref{fig:upperBound} (where all active UEs were replaced by their relays).  This shows that these low SINR UEs are Backhaul limited and thus almost exclusively scheduled through their relay. This is because these UEs (and their relays) tend to have poor pathloss to the base station and thus higher Access link transmit power (see Definition \ref{def:txPower} and subsequent discussion.)  Overall, the proposed design offers about $110\%$ increase in both the $5^{th}$ and $50^{th}$ percentile of average DL rate per UE, respectively.
  \item Average UL SINRs and rates also improve, though not as close to the upper bound as in the case of DL. Recall that UL upper bound ignores the Access-to-eNB interference (from both the edge-to-relay/DL and relay-to-edge/UL links.)
      Overall, the proposed design offers $240\%$ and $40\%$ increase in the $5^{th}$ and $50^{th}$ percentile of average UL rate per UE, respectively.
  \item Fig.~\ref{fig:intf} provides further insight into the UL results.  The figure shows the CDFs of various interferences and signal strengths at the base station antenna (capturing the fluctuation of these quantities from subframe to subframe.)  Keys are listed for curves from right to left.  We note that the gain in UL comes from both the reduction in ICI and increase in signal strength due the use of relays.  But while the ICI is reduced, Access-to-eNB interference is added due to the use of the relays (from both edge-to-relay/DL and relay-to-edge/UL links.)  Even though the CDF of Access-to-eNB interference is dominated by that of ICI, sometime it is indeed the Access-to-eNB interference that limits UL SINR.  As will be seen in next section, this Access-to-eNB interference is further weakened by the use of IM amongst Access links.
  \item Fig.~\ref{fig:acc} provides the CDF of average Access link SINRs, separately for the relay-to-edge (DL) and edge-to-relay (UL) direction.  As expected, edge-to-relay direction has better SINR since edge UEs have higher Access link transmit power (see Definition \ref{def:txPower} and subsequent discussion.) Note that Access links can achieve significant rates even at low SINRs due to reuse 1 of the entire UL spectrum on each link.  Links with poor average SINR can be those with low transmit power allowance, or those {\em surrounded} by UL UEs, or those co-located with another Access link (this factor will be mitigated by Access-to-access IM scheme presented in the next section.)
  \item Some other quantities of interest are as follows: $85\%$ of the DL UEs and $83\%$ of the UL ones are (at least partially) scheduled through relays while the rest are scheduled only directly.  On average $4.7$ Access links are simultaneously active per sector, as also reflected in approximately $13$dB separation between the Access-to-eNB interference CDF and signal strength CDF in Fig.~\ref{fig:intf}.
\end{enumerate}

\begin{figure}
\centering
\scalebox{1}{\includegraphics[trim = 7mm 0mm 10mm 0mm, clip, width=3.2in]{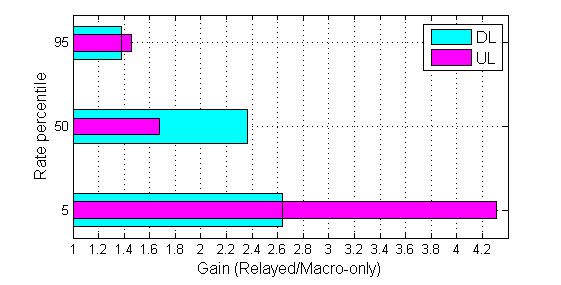}}
\caption{Improvement in DL and UL throughput, with interference management amongst Access links.}
\vspace{-0.2in}
\label{fig:gainIM}
\end{figure}

\subsubsection{With interference management amongst Access links}
With IM (\ie, $\gamma_{acc}=5$dB), Access links can support higher rate; Fig.~\ref{fig:acc} gives Access link SINR with and without IM.  Improved SINR can result in more traffic routed through relays (for bottle-neck Access links) or reduced Access-to-eNB interference (due to reduced 'on' duration).  The gains (over the legacy system) with IM are summarized in Fig.~\ref{fig:gainIM}.  We see $160\%$ and $135\%$ (vs $110\%$ without IM) increase in the $5^{th}$ and $50^{th}$ percentile of average DL rate per UE, respectively.  In case of UL, we see $330\%$ and $65\%$ (vs $240\%$ and $40\%$ without IM) increase in the $5^{th}$ and $50^{th}$ percentile, respectively.  This is due to the reduction in ICI and Access-to-eNB interference (shown in Fig.~\ref{fig:intf}) as a by-product of Access link IM.  There are now on average $2.5$ Access links simultaneously active per sector (vs $4.7$ links without IM).  ICI is also lower with IM; this is because more UL relays are being used ($94\%$ vs $83\%$) as well as more UL traffic now flows through relays than directly from the edge UE to eNB.  Number of DL UEs with relays has also increased to $96\%$ (vs $85\%$ without IM).

\section{Implementation and technology aspects}
\label{sec:implementation}

We discuss some of the signaling and implementation aspects mainly in the context of a 3GPP LTE system. In particular, we discuss protocols for relay discovery, scheduling of access links, and implications of power consumption.

\subsection{Signaling for relay selection}

Here, we present some protocols for power efficient discovery of idle mode UEs that can serve as a relay. The proposed design crucially depends on synchronization from the eNB and ability of idle UEs to measure downlink SINR without connecting to the eNB, both of which are supported in LTE.
As an example, consider $2$ UL sub-frames reserved for Relay discovery every second which amounts to $0.2\%$ of the resource. Amongst the 2 UL-subframes, PUSCH resources ($44$ RB-pairs per sub-frame) are reserved for relay discovery signaling. Each idle mode UE selects one out of the $88$ RB-pairs to transmit on, and transmits using rate $1/4$ QPSK code carrying 70 information bits. The information bits carry (i) UE identity (ii) UE's DL SINR (iii) access link information such as maximum transmit power on the access link. Note that $88$ resources are more than enough to avoid congestion as we typically see 20-30 candidate relays in the vicinity. So, even with spatial reuse of the relay discovery resource, a UE should be able to detect most of the candidate relay UEs. Additionally, certain optimizations can be done to stop UEs with low DL SINRs from participating in the relay discovery which can further reduce the congestion as well as the power penalty.

Now, an active mode UE that wants to use a relay receives on these two sub-frames and decodes up to $88$ relay discovery messages. Based on these messages it determines relays with highest DL SINR with certain restriction on the access link such as maximum path-loss. The active mode UE can select the best idle mode UE to relay its traffic, or alternatively the measurements can be reported to the eNodeB which can make the final relay association decision. We omit signaling for relay association from this paper.

\subsection{Signaling for interference management}

Due to introduction of access links on the uplink spectrum, some signaling is needed for managing interference between access links and uplink UEs.

{\em Interference from access link to the WAN} is managed through power control. For signaling, we propose to broadcast a single parameter indicating the maximum tolerable interference level from an access link to the base station -- this information could for example be carried on one of the system information broadcast (SIB) channels. It is possible to further optimize this, and broadcast the power cap per uplink assignment (as the received signal strength would vary based on the UE scheduled on the uplink). This would require the information being sent per assignment in PDCCH and can significantly increase the overhead.

{\em Interference from uplink UEs to access links} is managed based on statistical multiplexing as demonstrated in Section \ref{subsec:wan2accIM}, so no new signaling is needed.

{\em Interference between access links} can be managed through the eNB. In particular, it is proposed that UEs measure the periodically scheduled SRS signal from other UEs and determine UEs that can cause significant interference to a receiver and report that to the base station. The base station can then instruct access links to orthogonalize in time and/or frequency. This will be done through per UE dedicated signaling that would be carried on PUSCH/PDSCH. Alternatively, the interference can be managed in a distributed way as shown in \cite{WuTavildarShakkottai10}.

\subsection{Scheduling}
The Direct link or the Backhaul link can be thought of as a traditional LTE link and are scheduled as such.  For access link, the scheduling is implicit except for slow time scale scheduling if needed to manage access to access interference.

\subsection{Impact on power consumption}

One of the traditional concerns for opportunistic relaying is the power consumed at the relay UE. However, we look at the power consumption from a systems aspect, and argue that increase in power consumption is not significant. In particular, we consider:

{\em Probability of relaying: } given the large number of idle UEs in a network at a time, chance that a UE is selected to relay traffic for another UE is fairly small (about $4\%$ for the numbers used in this paper). Additionally, a UE with low battery can choose not to participate in the relay discovery protocol.

{\em Modem vs Device power consumption:} for a typical smartphone, modem is not the biggest power consumption (LCD backlight and application processor can consume comparable power \cite{CarrollHeiser10}). However, for a relaying UE only the modem needs to stay on thus the corresponding power drawn is much lower than what the device would consume in a typical scenario.

{\em No power amplifier over Access link:} since Access link transmissions are 3dBm or lower, the UE's power amplifier (PA) can operate in low power, high efficiency mode over Access link (see \cite[Fig.~7]{JensenLauridsen12}). As before, the high power low efficiency PA operation may be needed only at the device active on the cellular link.

{\em Power saving due to higher throughput:} power consumption in modem is typically dictated by the ON time -- however with $50\%$ median improvement in throughput, correspondingly power will be reduced by $50\%$ for given amount of bits exchanged with the network.

\section{Conclusion}
\label{sec:conclusion}
We presented a new architecture for two-hop relaying in cellular LTE networks, built on D2D communication.
The key idea proposed was to exploit shadow-diversity of idle UEs and reuse UL spectrum to relay both UL and DL traffic.
We argued that reusing UL spectrum facilitates easier interference management between Access and Backhaul links.
We validated the proposed architecture through simulations based on the 3GPP methodology, with proposed scheme showing 110\% gain in the median DL throughput and 40\% gain in the median UL throughput.
Finally, we presented a preliminary design for realizing the proposed scheme at a system level, including signaling for relay selection and interference management.
Throughout the paper we also pointed out various potential optimizations regarding relay selection, adaptive interference management, and scheduling, which will be investigated in future works.



\appendix
\label{sec:sysModel}
{\bf Channel model and UE drop:}
We borrow most of the details from Urban Case 1-3D,(10+10)MHz FDD scenario of the LTE evaluation methodology recommendations \cite[Annex A]{3GPP36814}.  We consider the standard two-tier, 19 cells with wrap-around deployments, inter-site distance (ISD) of 500 meters, and 3-sectored 3D eNB antenna pattern.  All UEs and eNBs have $1$ TX and $2$ RX antennas.  We assume 500 UEs per sector uniformly over space, out of which 20 are active -- 10 on the DL and the remaining 10 on the UL.  The distance-dependent pathloss between eNB and UE is given by $35.3+37.6\log_{10} d$, where the distance $d$ is in meters. Pathloss and shadowing in DL and UL spectrum are assumed to be identical. Shadowing standard deviation is set to $7$ dB and autocorrelation distance parameter $d_{corr}$ to $25$ meters.  Moreover, at any point, the shadowing values w.r.t different base stations have cross-correlation of $\frac{1}{2}$. A caveat about the (supposedly) shadowing correlation matrix $[\exp(\frac{d_{ij}}{d_{corr}})]_{ij}$ -- where $d_{ij}$ is the distance between the $i^{th}$ and $j^{th}$ UE {\em in the wrapped-around deployment} -- is that it is not guaranteed to be positive semi-definite.  However, with $d_{corr}$ much smaller than the diameter of the two-tier deployment, we numerically found the matrix to indeed be positive semi-definiteness.

UEs associate with the eNB with the best channel (lowest measured pathloss).  The UL transmit power of a UE with pathloss $PL$ dB and an allocation of $M$ resource blocks (RBs) is given by $\min\left(23\mbox{dBm},80\mbox{dBm} + 10\log_{10}M + 0.8PL\right)$.  SINR values for all links (UL, DL, and Access) are capped at $25$ dB. UE and eNB noise figures are $9$ dB and $5$ dB respectively, and noise power in $10$MHz bandwidth is $-104.5$ dBm.

The device-to-device pathloss model is taken from ITU-R P.1411-6 \cite{ITU1411}. More specifically, pathloss between two UEs at distance $d$ meters is given by,
$$
    PL_{d2d}(d) = \left\{
                            \begin{array}{ll}
                                38.47+20\log_{10}(d), & \mbox{if } d \leq 44; \\
                                44.85+40\log_{10}(d), & \mbox{if } d > 64; \\
                                71.34+2.29(d-44),     & \mbox{if } 44<d\leq 64. \\
                            \end{array}
                    \right.
$$
The third term -- pathloss between 44m and 64m -- is simply the linear interpolation between pathloss at 44m and 64m.

All traffic is assumed to be full-buffer.  The dynamics of the (small) buffer at the relay UE are naturally modeled, based on the activity of the Access and Backhaul links.

\end{document}